\documentstyle[prb,aps]{revtex}\tighten \twocolumn
\newif\ifepsf \global\epsffalse
\def\widetext{\par\global\columnwidth43.9pc
\global\hsize\columnwidth\global\linewidth\columnwidth
\global\displaywidth\columnwidth}
\widetext
\begin{document}
%%%%%%%%%%%%%%%%%%%%%%%%%%%%%%%%%%%%%%%%%%%%%%%%%%%%%%%%%%%%%%%%%%%%%%%%%%%
% This region is dedicated to definitions of simple macros that may be often
% used in the text
\def\ie{{\it i.e.}}
\def\fos{F_0^{s}}
\def\foa{F_0^{a}}
\def\fones{F_1^{s}}
\def\fonea{F_1^{a}}
\def\aos{A_0^{s}}
\def\aoa{A_0^{a}}
\def\aosa{A_0^{s\!,a}}
\def\alsa{A_\ell^{s\!,a}}
\def\fosa{F_0^{s\!,a}}
\def\flsa{F_\ell^{s\!,a}}
\def\fls{F_\ell^{s}}
\def\fla{F_\ell^{a}}
\def\SLTO{Sr${}_{1-x}$La${}_x$TiO${}_3$}
\def\LTO{LaTiO${}_3$}
\def\STO{SrTiO${}_3$}
\def\half{{1\over2}}
\def\WR{R_w}
\def\pp{p^\prime}
\def\ppp{p^{\prime\prime}}
\def\bp{{\bf p}}
\def\bq{{\bf q}}
\def\bpp{{\bf p}^{\prime}}
\def\GIR{\Gamma^{{\rm IR}}}
%END of defs!
%%%%%%%%%%%%%%%%%%%%%%%%%%%%%%%%%%%%%%%%%%%%%%%%%%%%%%%%%%%%%%%%%%%%%%%%%%%
\title{ Robustness of a local Fermi Liquid against Ferromagnetism and
        Phase Separation }
\author{
        Jan R. Engelbrecht${}^{1,2}$
        and
        Kevin S. Bedell${}^{2}$
        }
\address{
              ${}^1$Center for Materials Science MS-K765,
              Los Alamos National Laboratory,
              Los Alamos, NM 87545
\\
              ${}^2$Theoretical Division MS-B262,
              Los Alamos National Laboratory,
              Los Alamos, NM 87545
              }
\maketitle
\begin{abstract}
We study the properties of Fermi Liquids with
the microscopic constraint of a local self-energy.
In this case the forward scattering sum-rule imposes strong limitations on the
Fermi-Liquid parameters, which rule out any Pomeranchek instabilities.
For both attractive and repulsive interactions, ferromagnetism and phase
separation are suppressed.  Superconductivity is possible in an $s$-wave
channel only.
We also study the approach to the metal-insulator transition, and find
a Wilson ratio approaching 2.
This ratio and other properties of \SLTO\ are all
consistent with the local Fermi Liquid scenario.
\end{abstract}

\vspace{-35pt}
\par\vspace{+30pt}

%%%%%%%%%%%%%%%%%%%%%%%%%%%%%%%%%%%%%%%%%%%%%%%%%%%%%%%%%%%%%%%%%%%%%%%%%%%%

For more than a decade now there has been considerable interest in
understanding the strong electronic correlations in a variety of
materials such as the heavy-fermion metals and related insulators
and
the perovskites such as the high-$T_c$ cuprate systems.
Recurrent themes in the study of these materials are issues such as
unusual Fermi Liquid states characterised by large effective masses, and
small
Wilson ratios, the proximity to metal-insulator transitions
or instabilities towards ferromagnetism, anti-ferromagnetism or
superconductivity.
Within various frameworks, and to varying degrees of approximation,
these systems have been modeled by theories of electrons with
local (\ie\ momentum-independent) self-energies.
For instance, there is much interest in
studying correlated systems such as the Hubbard Model using
dynamic mean-field
theories\cite{inf-D} that become exact in infinite dimension,
(based on describing electronic correlations through a local self-energy).

With this background, we investigate the general consequences
for Fermi-Liquid Theory, given the microscopic constraint of a
local self-energy (arising from quasiparticle correlations alone).
We shall show that our local Fermi Liquid (LFL) can never satisfy the
Stoner criterion and is robust against instabilities towards ferromagnetism
as well as phase separation.
This result has strong implications for the use of dynamic mean-field
theories to models describing Fermi Liquids.
Particularly in the context of recent studies\cite{no-ferro} into the
criteria necessary for ferromagnetism in metals
(due to electronic interactions)
as well as problems with phase separation in superconducting models.

In addition, we investigate the approach to the metal-insulator transition
(from the metalic side) for our LFL and find a Brinkman-Rice transition
with a Wilson ratio, $\WR$, near 2 for a wide range of parameters.
These results compare very well with measurements\cite{SLTOa}
of $\WR$, the spin susceptibility $\chi_s$ and the effective mass $m^{*}/m$
in the perovskite \SLTO\ which becomes insulating as the doping $x\to1$.
Furthermore, out LFL scenario makes several predictions which can be
measured to test the validity of characterising this system as a LFL.
We also investigate the Ward identities and vertex corrections in our theory
and contrast our self-consistent LFL with descriptions of
Fermi Liquids based on the Gutzwiller approximation which includes some
aspects of the local self-energy without being self-consistent.
Finally we discuss the limitations of our theory, especially as far as
anti-ferromagnetic correlations go.  We also briefly outline a natural
generalisation of our approach which allows the incorporation
of anti-ferromagnetic fluctuations in a self-consistent way.

Our starting point is to take as given that we have a
single-band\cite{foot-band}
Fermi Liquid with a momentum-independent self-energy
$\Sigma(\omega)$, where $Im\,\Sigma(\omega)\ll\omega$ as $\omega\to0$.
Further, we consider the simplest case where the interactions are
mediated by the quasiparticles themselves.
We emphasise that we do {\it not} make the assumption of infinite dimension,
although a Fermi Liquid in infinite D would necessarily be a realisation
of the LFL theory we present here.

%%%%%%%%%%%%%%%%%%%%%%%%%%%%%%%%%%%%%%%%%%%%%%%%%%%%%%%%%%%%%%%%%%%%%%%%%%%%

In the spirit of Landau's microscopic foundation of Fermi-Liquid Theory we
determine the structure of the 4-point vertex,
given\cite{agd} by
\begin{equation}\label{eqn:vert}
  \Gamma_{p,\pp\!}(q)
  \! = \!
  \GIR_{p,\pp\!}(q)
  \!-\!
   i \!\!\! \int_{\ppp} \!\!\! \GIR_{p,\ppp}\!(q)
     G(\ppp\!\!+\!q) G(\ppp) \Gamma_{\ppp\!\!,\pp\!}(q)
\end{equation}
where $\GIR$ is the irreducible particle-hole vertex,
$G$ is the exact Green's function and we have dropped the spin indices.
Here we introduce the notation
$p=(\bp,\omega)$,
while
$q=(\bq,\Omega)$ denotes the momentum/energy transfer and
$\int_{\ppp}$ represents the (D+1)-dimensional integral with appropriate
measure.

Using the standard analysis, $\Gamma$ can then be identified\cite{agd}
with the quasiparticle interaction
in the appropriate small momentum transfer limit,
with for $\bp$ and $\bpp$ on the Fermi surface
and quasiparticle energies $\omega$, $\omega^\prime=0$,
\begin{equation}\label{eqn:fpp}
z^2\,f(\bp,\bpp)=\lim_{\Omega\to0}\lim_{|\bq|\to0}\Gamma_{p,\pp}(q)
{}.
\end{equation}
Introducing the Fermi-Liquid parameters and the scattering amplitudes in the
standard way, (\ref{eqn:vert}) then leads to the usual relation
\begin{equation}\label{eqn:a:vs:f}
  \alsa={\flsa\over1+\flsa/(2\ell+1)}
\end{equation}
where
$N(0)\,f_{\sigma,\sigma^\prime}(\bp,\bpp)
=
\sum_l(\fls+\vec\sigma\cdot\vec\sigma^\prime\fla)P_\ell(\hat\bp\cdot\hat\bpp)$
and $N(0)$ is the density of states.

Consider the general case of a conserving, $\Phi$-derivable theory
where
the relations between $\GIR$, $G$ and the self-energy $\Sigma$ are given by
%\begin{equation}
$
  \Sigma(p)={\delta\Phi/\delta G(p)}
$
%\end{equation}
and
%\begin{equation}
$
  \GIR_{p,\pp}=(i/2){\delta^2\Phi/\delta G(\pp)\delta G(p)}
{}.
$
%\end{equation}
Integrating the second-order functional derivative then yields
\begin{equation}\label{variation-eqn}
  \delta\Sigma(p)=-(i/2)\int_{\pp}\GIR_{p,\pp}(0)\;\,\delta G(\pp)
{}.
\end{equation}
In order to understand the constraints imposed by a local self-energy we
simply take $\Sigma$ in this equation to be momentum-independent.
Together with the symmetry\cite{foot-symmetry} of $\GIR$
under $p\!\!\leftrightarrow\!\!\pp$,
this immediately implies that the irreducible vertex is also local, \ie,
$\GIR_{p,\pp}(0)
 \to\GIR_{\omega,\omega^\prime}(0)
 =\delta\Sigma(\omega)/\delta G_{\rm loc}(\omega^\prime)
$
with
$ G_{\rm loc}(\omega)=\sum_\bp G(\bp,\omega)$.

The local irreducible vertex is analytic in $q$ for small $q$ and hence
$\GIR_{\omega,\omega^\prime}(q)$ is also local which
in turn implies a local full vertex, $\Gamma_{\omega,\omega^\prime}(q)$.
It then follows from (\ref{eqn:fpp})
that $f(\bp,\bpp)$ is independent of the angles between the momenta,
and consequently
only $s$-wave Fermi-Liquid parameters and scattering amplitudes can
be non-zero.
Moreover, the forward-scattering sum-rule
$ \sum_\ell \left(A_\ell^{s}+A_\ell^{a}\right)=0 $ now simplifies to
\begin{equation}\label{eqn:sumrule}
  \aoa=-\aos
\end{equation}
with $|\aosa|\le1$ and (\ref{eqn:a:vs:f}) then implies
\begin{equation}\label{eqn:f:rel}
  \foa=-\fos/(1+2\fos)
\end{equation}
with $-\half\le\foa,\fos<\infty$.
This relation is plotted in Fig~1 for the repulsive case ($\fos>0$).
For the relatively large $\fos$ expected for a
strongly correlated local Fermi Liquid, $\foa$ very rapidly saturates
to $-\half$.
For comparison we also show the results of M\"uller-Hartmann's
second-order perturbative calculation \cite{FLP-infD}
of the Fermi-Liquid parameters of the Hubbard model in infinite D.
\ifepsf%
\par\vspace{-2.4cm}\par
\begin{figure}[bth]
 {\epsfxsize=4.00in\epsfbox[298 354 623 608]{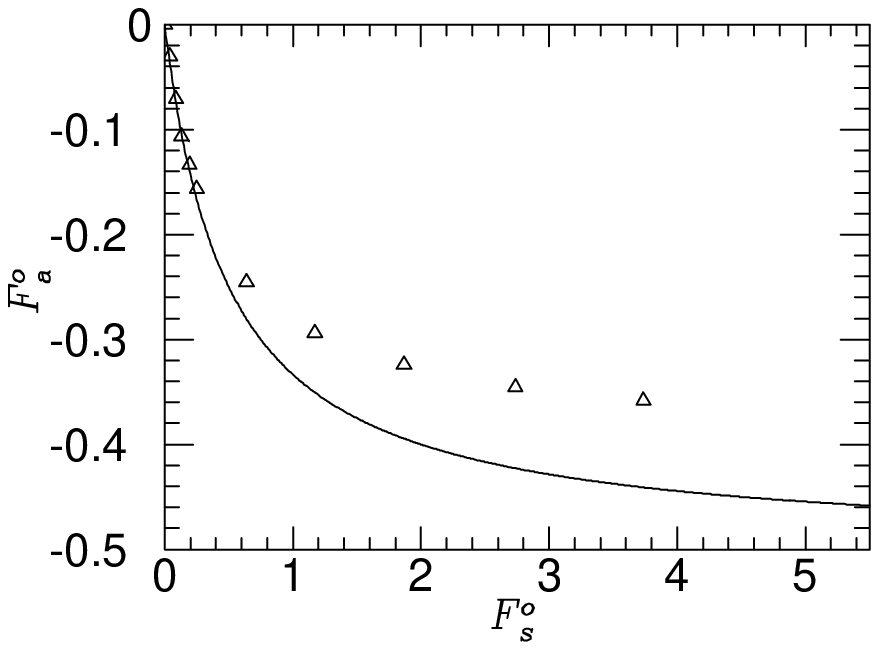}}
  \caption{The relation between $\fos$ and $\foa$ in (6)     %(\ref{eqn:f:rel})
        -- the circles are for the Hubbard model$^7$         %\cite{FLP-infD}
           in infinite D to $2^{\rm nd}$ order in $U$.
           \label{fig:fo's}}
\end{figure}
\else\par\fi
An additional simplification for the local Fermi Liquid is that
since $\partial\Sigma/\partial\bp=0$, the
quasiparticle residue and the effective mass
are rather trivially related:
\begin{equation}
    z^{-1} = m^{*}/m
{}.
\end{equation}

The microscopic constraint of a local self-energy,
is clearly non-trivial.
Whereas Fermi-Liquid Theory is characterised in general by an infinite
number of Fermi-Liquid parameters,
the local Fermi Liquid
is quite generally described by
only {\it two} independent parameters, for instance, $m^{*}$ and $\fos$.
The resulting two-parameter phenomenology then implies strong relations
between various experimental measurements which can be easily tested
to verify the applicability of the LFL scenario.
These results are for the case of quasiparticle mediated interactions
which lead to (\ref{eqn:vert}).
In the case of electron-phonon\cite{electron-phonon1,electron-phonon2}
and single impurity Kondo\cite{1imp-kondo}
Local Fermi Liquids the interactions are mediated through exchange of phonons
or excitations of the impurity-electron singlet which would modify the form of
(\ref{eqn:vert}) and hence the relations between the parameters.
Finally, we note that
the momentum independence of $\Sigma$, also trivially implies that
not only the size but also the shape of the Fermi surface are not changed by
interactions.

%%%%%%%%%%%%%%%%%%%%%%%%%%%%%%%%%%%%%%%%%%%%%%%%%%%%%%%%%%%%%%%%%%%%%%%%%%%%

Since we start off with a $\Phi$-derivable theory, our description is
conserving by construction and satisfies the Ward identities.
The analysis of the 3-point vertices for the charge and spin densities
again starts from (\ref{variation-eqn})
and yields that these vertices are also local,
for small energy/momentum transfers,
\ie, $\Lambda^{(q)}(p)\to\Lambda^{(q)}(\omega)$.
Setting $\omega=0$, we consider the $|\bq|,\Omega\to0$ limit: $\Lambda^r$
with $r=|\bq|/\Omega$.
Using the Ward identities one then gets\cite{agd},
\begin{equation}
     z\Lambda^0       =     1
\end{equation}
\par\vspace{-5pt}\par
\noindent and
\par\vspace{-5pt}\par
\begin{equation}\label{eqn:dens:vert}
     z\Lambda^\infty  =    1/(1+\fosa)
\end{equation}
where $\fos$ ($\foa$) applies for the charge (spin) density.

The 3-point vertices for the charge and spin currents do retain a trivial
momentum dependence, but are otherwise completely unrenormalised (in the
interesting two limits\cite{foot-current-vert}).
For both spin and charge, in the direction $\hat e_\alpha$, we get
\begin{equation}
     z\Lambda_\alpha^0       =     p_\alpha/m^{*}
\end{equation}
\par\vspace{-5pt}\par
\noindent and
\par\vspace{-5pt}\par
\begin{equation}
     z\Lambda_\alpha^\infty  =     p_\alpha/m^{*}
{}.
\end{equation}
{}From these relations it can be shown that
there are no vertex corrections to the optical conductivity.
Without the local assumption, these relations normally do
depend\cite{current-vert}
on Fermi-Liquid parameters.

We emphasise that all our results hold for arbitrary dimension --- provided
one has a
single-band\cite{foot-band}
Fermi Liquid, all that was required was a local self-energy.
The dynamical mean-field theories which become exact in infinite dimensions,
also have local vertex corrections and unrenormalised current vertices, but
our results indicate that
(at least in the small $q$ limit) this is a consequence of the local
self-energy, rather than infinite D, per se.

%%%%%%%%%%%%%%%%%%%%%%%%%%%%%%%%%%%%%%%%%%%%%%%%%%%%%%%%%%%%%%%%%%%%%%%%%%%%

Now we are in a position to analyse the possible instabilities of our local
Fermi Liquid.
{}From (\ref{eqn:dens:vert}) the spin susceptibility $\chi_s=N(0)/(1+\foa)$
while the compressibility $\kappa=n^2\,N(0)/(1+\fos)$.

Consider an interaction which is repulsive in the singlet channel, in which
case $\fos>0$ and $-\half<\foa<0$.
The Stoner criterion, $1+\foa=0$, can then never be met
and hence the LFL is stable
against a ferromagnetic instability.
Further, since eqn. (\ref{eqn:vert}) cannot generate interactions in higher
angular-momentum channels, the LFL is also robust against the Kohn-Luttinger
type superconductivity.
(Also, phase separation is ruled out since $\fos>0$.)
The only potential instability is towards anti-ferromagnetism and/or
a metal-insulator transition (see below).

For an attractive interaction, one has $-\half<\fos<0$ and $\foa>0$ and
there can then be no phase separation since the Pomeranchek instability
signaled by the condition $1+\fos=0$ can also never be satisfied.
(Additionally, ferromagnetism is ruled out since $\foa>0$.)
The only possible instability would be towards the BCS state.

%%%%%%%%%%%%%%%%%%%%%%%%%%%%%%%%%%%%%%%%%%%%%%%%%%%%%%%%%%%%%%%%%%%%%%%%%%%%

The approach to the metal-insulator transition (from the metalic side) is
signaled by a vanishing charge Drude weight $D=ne^2/m^{*}$.
Recall that the Fermi surface volume is unrenormalised by correlations,
so the only possible way $D$ can vanish in our framework, is for the
effective mass to diverge.  The metal-insulator transition is then
Brinkman-Rice like.
A diverging $m^{*}$ would imply a diverging density of states, and
(although $\fos$ is an independent parameter)
it is quite natural to expect $\fos\sim N(0)$ to become quite large in this
limit.
It then follows from (\ref{eqn:f:rel}) that $\foa$ would rapidly approach
its limiting value of $-\half$.
This corresponds to the unitarity limit
with the scattering amplitude $\aos\to1$.
Within this scenario, the Wilson ratio
$\WR=\chi_s/N(0)=1/(1+\foa)$
should rapidly saturate at 2.
For a strongly correlated system one expects $\fos>1$ in which case $\WR$
would also be close to 2.
In the case of an attractive interaction, $\foa$ is necessarily positive
and then one would find\cite{foot-heavy} $\WR<1$.

The LFL theory, developed above, seems to apply particularly well to
the perovskite system \SLTO,
which exhibits a metal-insulator transition upon doping.
In the $x=0$ limit, \STO\ is a band insulator
for which band calculations\cite{band-struct}
predicts a band mass $m_b\sim2m_e$.
On the other hand for $x=1$ (corresponding to the $\half$-filled case)
\LTO\ is an insulating anti-ferromagnet.
Recent experiments\cite{SLTOa,SLTOb} have carefully elucidated the doping
dependence in the intermediate metalic regime
$0.3<x<0.95$, and have shown that this material is very well
described by a Fermi Liquid state with a Brinkman-Rice transition as $x\to1$.
Specifically, this is borne out by measurements on the resistivity:
$\rho(T)-\rho_o\sim AT^2$ and the specific heat:
$C_V\sim\gamma T$
which show typical Fermi Liquid behaviour for $0.3<x<0.95$.
Also, the Hall coefficient: $1/R_H\sim x$,
is electron-like and temperature independent.
Upon increasing $x$, the effective mass diverges
and $\gamma$, $\chi_s$, the inverse Drude weight and
the inverse of the square of the plasma frequency,
which are all proportional to $m^{*}$, increase rapidly.
However, even though both $\gamma$ and $\chi_s$ have a strong
doping-dependence, the Wilson ratio $\WR\propto\chi_s/\gamma$
is essentially doping-independent
and saturates at a value of 2.
{}From the low-frequency optical conductivity measurements\cite{SLTOb},
the effective mass at $x=0.5$ is about $m^{*}\approx5m_e$ and increases
rapidly with increasing $x$.

The resulting picture of \SLTO\ is that the electronic response
is that of a simple, yet strongly-correlated, single-component Fermi Liquid.
This state is strongly correlated in the sense that $m^{*}$ and hence
$\fos$ are large and that the system exhibits a metal-insulator
transition.
Yet the Fermi Liquid state is very simple, in that for all doping,
it seems to be well described in terms of a single parameter namely,
$m^{*}(x)$,
with the only relevant scale being the Fermi energy
$\epsilon_F=k^2_F/2m^{*}(x)$.
(Also, $\foa$ seems to be pinned at $-\half$.)
This is reminiscent of the single impurity Kondo problem where strong
interaction effects lead to a rather simple low-energy Fermi Liquid state.

All the properties of this material
corresponds remarkably well with our local Fermi-Liquid Theory
described above.
Indeed, if we take a simple parabolic band with
$\fos=c_1m^{*}(x){\root3\of{x}}$,
using the experimental value of $\gamma$ to scale the mass in this formula,
the resulting theoretical estimation of $\WR$ agrees quite well with
experiment.
In Fig.~3 we plot the doping dependence of the Wilson ratio for
two plausible choices of $c_1$,
with the actual experimental results plotted as the triangles.
In principle, allowing the coefficient $c_1$ to depend on $x$, would do better,
but the main point is that this simple picture captures the essence of the
behaviour of $\WR$.
With $\WR$ close to 2 it is not surprising that superconductivity has not been
observed in \SLTO.  If it were found to be superconducting at very low
temperature that would require some small departure from LFL behaviour.
\ifepsf%
\par\vspace{-2.2cm}\par
\begin{figure}[bth]
 {\epsfxsize=4.00in\epsfbox[298 334 623 548]{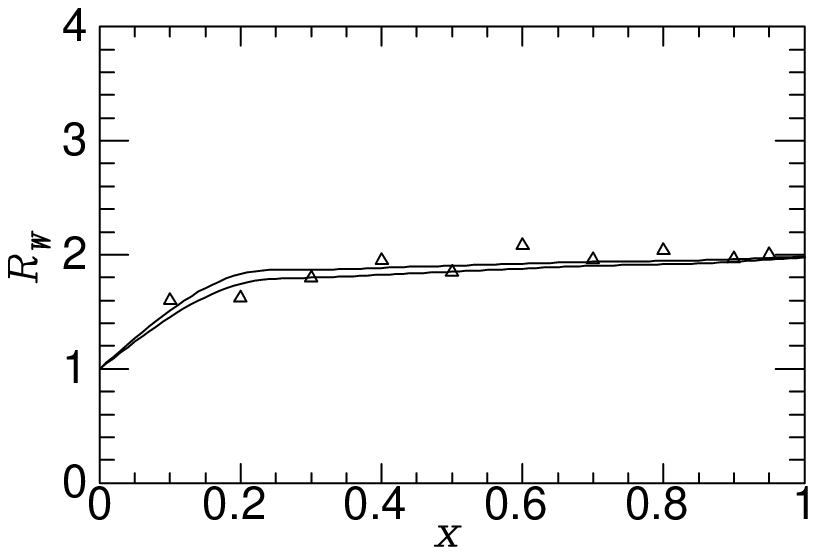}}
  \caption{Wilson ratio, $\WR$ {\it vs.} doping $x$
           the triangles are the experimental values from ref.~3. %\cite{SLTOa}
           \label{fig:wilson-ratio2}}
\end{figure}
\fi

One of the quantities we can estimate is the coefficient of the $T^2$ term in
the resistivity.
The general form for this term
has been evaluated\cite{lawrence-wilkins}
to be
\begin{equation}
\rho(T)=\rho_o
+\left(
 {3\pi\over16}{\langle|W|^2\rangle\over p_F a_B n^2} \gamma^2
 \right)
 T^2,
\end{equation}
where the coefficient in parentheses is denoted by $A$.
For the local Fermi Liquid the scattering amplitude simplifies to
$\langle|W|^2\rangle=8(\aos)^2$
where we have replaced the transport lifetime by the quasiparticle lifetime.
This will usually give an order of magnitude estimate for $A$
and at times it works even better\cite{lawrence-wilkins,bedell}.
In general, $A$ is not strictly proportional to $\gamma^2$,
however, near the metal-insulator transition our model predicts that
$\langle|W|^2\rangle$
saturates at a value of 8 and, apart from the small change in $n$,
the dominant change is through $\gamma^2$.
This is clearly observed in the data for \SLTO\ with $x$ close to one.
In fact, the simple parabolic band picture gives an estimation of $A$
which agrees with the experiments up to a factor of two.
It should be emphasised that, in principle, $A$ could be completely dominated
by the scattering amplitude --- particularly close to an instability.
The fact that it is not is a non-trivial vindication of our claim
that \SLTO\ is a local Fermi Liquid.

Another prediction to follow from this theory is the coefficient of the
$T^3\ln T$ term in the specific heat.
This is given by\cite{baym-pethick},
\begin{equation}
C_V
=\gamma T+
\left(
   {3\pi^2 \gamma (\aoa)^2 \over 5 \epsilon_F^2}
   \left(1-{\pi^2\aoa\over24}\right)
\right)
T^3\ln(T/\epsilon_c)
\end{equation}
where we have employed the simplifications for a LFL.
Near the metal-insulator transition
the coefficient in parentheses should saturate to
about $8.4\gamma/\epsilon_F^2$
near the metal-insulator transition.
This prediction could be tested by plotting
$ (C_V/T-\gamma)/T^2 $
{\it vs.\/}
$ \ln T $
and determining the slope.

A quantity related to the specific heat is the thermopower
$ Q=\gamma T / 3ne $.
Since the carriers are electron-like ($e\!<0\!$), $Q$ should be negative.
With
$k_B/2e\sim-43\mu V/K$
and
$ \epsilon_F\sim4\times10^3\;K$
($x\!=\!0.6$)
this gives
$ Q\sim-0.04 (\mu V/K^2) T$.
While getting the numbers right on the thermopower depends on many factors,
the trends should be right from our picture.
Thus,
$Q/T$ will grow as $\gamma/n$ as $x\to1$.

In this model we have shown that $\flsa\!=\!\!0$ for $\ell\!\ge\!1$.
While a direct measurement of $\fones$ is not possible,
a conduction-electron spin-resonance (CESR) experiment could determine the
value of $\fonea$.
In such an experiment the spin waves have a dispersion\cite{CESR},
\begin{equation}\label{eqn:CESR}
\omega_q
=
\omega_L
+
\left({1\over3}-{1+\foa\over3+\fonea}\right)
{q^2v_F^2\over\omega_L}
\end{equation}
with $\omega_L$ the Larmor frequency.
In a LFL, the
coefficient in parenthesis
should reduce to $-\foa/3$.
Finding such a correspondence between $\omega_q$ and $\foa$ as determined
from $\WR$ would clearly be the most direct evidence for a LFL.

As $m^{*}$ increases close to the metal-insulator transition,
the only energy scale in the LFL namely, $\epsilon_F$ reduces accordingly.
This effect, as well as the building anti-ferromagnetic correlations,
should lead to the breakdown of the LFL framework
sufficiently close to the transition.
For \SLTO\ the Fermi Liquid state is surprisingly robust
and anti-ferromagnetic order sets in only at $x\sim0.99$.
What happens beyond $x=0.99$?
Fermi-Liquid Theory, per se, focuses on small momentum-transfer processes and
thus does not address the physics of anti-ferromagnetism.
In general, keeping track
of $(\pi,\pi,\ldots)$-scattering processes associated with anti-ferromagnetism
is a rather formidable task.
However, a
generalisation of our approach which allows the
incorporation of anti-ferromagnetic fluctuations is to introduce two
(bi-partite) sublattices and allow different self-energies on each sublattice
while enforcing the constraint of a local self-energy on each sublattice.

It is insightful to
compare our results for the LFL with those obtained using the
Gutzwiller approximation.
This approximation,
as used by Brinkman and Rice\cite{brinkman-rice}
and discussed in detail by Vollhardt\cite{vollhardt-he3},
sets $m^{*}/m=z^{-1}$.
The quasiparticle residue
$z$ is determined from the theory and the assumption is made
that the self-energy is local, \ie,
$\delta\Sigma/\delta{\bf p}=0$ in order to determine $m^{*}$.
It was shown\cite{vollhardt-he3}
that the Wilson ratio saturates at 4 when this method
is applied to a Galilean invariant system
such as liquid ${}^3$He where $\fones\ne0$.
This is clearly in contradiction with the LFL notions we have presented here.
The reason is that, with the exception of $D=\infty$,
the Gutzwiller approximation is {\it not} local as it is assumed to be.
Whereas the Gutzwiller approximation is close in spirit to the LFL ideas
($\WR$ does saturate at high pressure
and ferromagnetism is suppressed), it is not
a proper local Fermi Liquid.
In fact, as we have shown, the consequence of a local self-energy for
a Galilean invariant system is that $m^{*}/m=z^{-1}=1+\fones/3=1$
which is clearly not the case in ${}^3$He.
The Gutzwiller approximation scheme has been recently applied\cite{gutzwiller}
to \SLTO\ but that calculation again suffers from these subtle inconsistencies
regarding locality and would fail to explain the Wilson ratio.
A local Fermi Liquid scenario
based on the Nozi\`eres' single impurity Kondo Fermi Liquid\cite{1imp-kondo}
was introduced by Varma {\it et. al.}\cite{varma}
for the heavy-fermion compounds.
This is a special case of our LFL theory with $m^{*}/m=1+\fos$.
This relationship for $m^{*}$ does not hold in general, but
follows from an additional assumption of an unrenormalised compressibility.

%%%%%%%%%%%%%%%%%%%%%%%%%%%%%%%%%%%%%%%%%%%%%%%%%%%%%%%%%%%%%%%%%%%%%%%%%%%%

In conclusion,
we have developed the self-consistent theory of a Fermi Liquid with the
microscopic constraint of a local self-energy.
We have shown that the local Fermi Liquid is
robust against instabilities towards
ferromagnetism, phase separation and Kohn-Luttinger superconductivity.
The metal-insulator transition in the LFL is
Brinkman-Rice-like, and is associated with a Wilson ratio of 2.
The properties of our LFL seems to describe the material \SLTO\
quite well, and we have made several predictions for this material.
Our results also have important consequences for dynamic mean-field theories
used in infinite D especially when considering states in which all
quasiparticles are equivalent -- in contrast to anti-ferromagnetic states.

%%%%%%%%%%%%%%%%%%%%%%%%%%%%%%%%%%%%%%%%%%%%%%%%%%%%%%%%%%%%%%%%%%%%%%%%%%%%

We would like to acknowledge fruitful discussions with
A. Balatsky, G, Kotliar, A. Ruckenstein, Q, Si, S. Trugman and D. Vollhardt
as well support of the DOE and the warm hospitality of the Aspen Center for
Physics where a significant part of this work was done.
K.B. would also like to acknowledge support from
the Institute of Advanced Studies in Canberra and the hospitality of M.P. Das.

\par\vspace{-0.5cm}\par

%%%%%%%%%%%%%%%%%%%%%%%%%%%%%%%%%%%%%%%%%%%%%%%%%%%%%%%%%%%%%%%%%%%%%%%%%%%%

%%%%%%%%%%%%%%%%%%%%%%%%%%%%%%%%%%%%%%%%%%%%%%%%%%%%%%%%%%%%%%%%%%%%%%%%%%%%
%%               FIGURES
\ifepsf%
\else
\begin{figure}[bth]
  \caption{The relation between $\fos$ and $\foa$ in (6),    %(\ref{eqn:f:rel})
           the circles are for the Hubbard model$^7$         %\cite{FLP-infD}
           in infinite D to $2^{\rm nd}$ order in $U$.
           \label{fig:fo's}}
\end{figure}

\begin{figure}[bth]
  \caption{Wilson ratio, $\WR$ {\it vs.} doping $x$
           the triangles are the experimental values from ref.~3. %\cite{SLTOa}
           \label{fig:wilson-ratio2}}
\end{figure}
\fi

\begin{references}
\bibitem{foot-band}
By single band we mean there is one species of fermion, not only
in the hamiltonian but also in the ground state.
\bibitem{foot-current-vert}
There is renormalisation of current vertices for $0<r<\infty$.
\bibitem{foot-heavy}
Non-superconducting heavy-fermion materials typically have
$1\!<\!\WR\!<\!2$ while $\WR\!<\!1$ for their superconducting cousins.
\end{references}
\end{document}